\documentclass[prl,reprint]{revtex4-2}

\usepackage{appendix}
\usepackage{amsfonts,bm}
\usepackage{amsmath,amssymb,amsthm}

\usepackage{graphicx}
\usepackage{epstopdf}
\usepackage{dcolumn}
\usepackage{mathrsfs}
\usepackage{tikz}
\usetikzlibrary{positioning}
\usepackage[colorlinks=true,linkcolor=blue,citecolor=blue, urlcolor=blue,bookmarks=false]{hyperref}
\usepackage{qcircuit}

\usepackage{tikz}	
\usetikzlibrary{backgrounds,fit,decorations.pathreplacing}

\begin{document}
\newtheorem{theorem}{Theorem}[section] 
\newtheorem{lemma}[theorem]{Lemma}

\title{Theory of Order-Disorder Phase Transitions Induced by Fluctuations Based on Network Models}
\date{\today}

\author{Yonglong Ding}
\affiliation{Beijing Computational Science Research Center, Beijing 100193, China}
\affiliation{School of Physics, Beijing Institute of Technology, Beijing 100081, China}

\begin{abstract}
Both quantum phase transitions and thermodynamic phase transitions are probably induced by fluctuations, yet the specific mechanism through which fluctuations cause phase transitions remains unclear in existing theories. This paper summarizes different phases into combinations of three types of network structures based on lattice models transformed into network models. These three network structures correspond to ordered, boundary, and disordered conditions, respectively. By utilizing the transformation relationships satisfied by these three network structures and classical probability, this work derive the high-order detailed balance relationships satisfied by strongly correlated systems. Using the high-order detailed balance formula, this work obtain the weights of the maximum entropy network structures in general cases. Consequently, I clearly describe the process of ordered-disordered phase transitions based on fluctuations and provide critical exponents and phase transition points. Finally, I verify this theory using the Ising model in different dimensions, the frustration scenario of the triangular lattice antiferromagnetic Ising model, and the expectation of ground-state energy in the two-dimensional Edwards-Anderson model.
\end{abstract}

\maketitle
\section{Introduction}

Through advancements in computational methods and models tailored for phase transitions, our understanding of these phenomena has become increasingly profound and comprehensive. However, existing algorithms such as Monte Carlo\cite{luo2022hybrid,prokof2008fermi,RevModPhys.73.33,moutenet2018determinant,rubtsov2005continuous}, tensor networks\cite{orus2019tensor,gray2021hyper,yuan2021quantum,pan2022simulation,hu2019dirac}, and scaling laws\cite{shao2016quantum,halperin2019theory,manipatruni2019scalable} all confront significant challenges, namely, a steep increase in computational complexity as the system size grows\cite{ortega2015fpga,yang2019high,preis2009gpu,meredith2009accuracy}. 
Lars Onsager\cite{PhysRev.65.117} and Chen-Ning Yang \cite{yang1952spontaneous} determined the phase transition point and critical exponent $\beta$ for the two-dimensional Ising model. However, an analytical solution for the three-dimensional Ising model has not yet been obtained. Currently, the most accurate value derived through numerical algorithms for the critical exponent $\beta$ is 0.32641871(75)\cite{PhysRevE.97.043301,PhysRevE.108.034118}.
Consequently, there remains a lack of comprehensive understanding of phase transitions at infinite scales.

Both quantum phase transitions \cite{vojta2003quantum} and thermodynamic phase transitions \cite{santen2000absence} can potentially be induced by fluctuations \cite{nelson2020quantum,mishin2015thermodynamic}. However, how fluctuations specifically cause phase transitions remains an area where Landau's theory of phase transitions provides only a phenomenological explanation \cite{hohenberg2015introduction}, lacking a clear description of the microscopic mechanisms underlying phase transitions. 

Ding\cite{Ding202401,Ding202402} recently developed a computational method that directly converts infinite-scale models into network models with finite nodes and constructed a maximum entropy network structure\cite{PhysRevX.14.041051}. This method, like the Bortz-Kalos-Lebowitz algorithm\cite{BORTZ197510}, begins with the classification of lattice points but employs a distinct classification scheme. This approach allows for the estimation of critical points and critical exponents $\beta$ for the Ising model's phase transition. In this paper, I comprehensively enhance this method by proposing a high-order detailed balance relationship based on network models. Taking the maximum entropy network structure as the starting point and building a bridge between fluctuations and phase transitions, this method provides a clear picture of how fluctuations induce phase transitions and offers an estimation method for critical points and critical exponents in ordered-disordered phase transitions under general conditions. 
It is worth noting that this computational method constructs network models based on fluctuations without distinguishing between thermodynamic and quantum phase transitions, making it applicable to both types of phase transitions. This paper only considers the case of a single lattice site with a single spin; subsequent explorations will address scenarios involving vacancies, double occupations, and even exchanges, which, due to the Heisenberg uncertainty principle, will become more complex and distinct from the present case. To validate the correctness and advantages of this theory, in addition to deriving equations for the variation of magnetic induction intensity with temperature in different-dimensional Ising models,
I also present frustration scenarios in antiferromagnetic Ising models\cite{liao2017gapless,hu2019dirac,ferrari2019dynamical} and estimates of ground-state energy in Edwards-Anderson models\cite{dominguez2011characterizing}.

\section{Theory} 
For lattice models, the infinite lattice model is transformed into a network model. Firstly, all possible lattice sites in the lattice model are classified according to the magnitude of their interactions and the spin of the lattice sites themselves. Let $C_{ij}$ represent the weights of different types of lattice sites, where $i$ denotes the spin type of the lattice site itself, and $j$ represents different types of neighbor interactions. Then, different $C_{ij}$ values are treated as different network nodes. If there exists a transformation relationship between different network nodes caused by fluctuations, the two network nodes are connected by a line segment. Finally, the different phases of the lattice model are labeled using the weights of different network nodes. If all lattice sites have spins pointing upwards, the weight of the corresponding network node is set to $1$, while the weights of other network nodes are set to $0$. This method does not focus on the specific position and momentum of any individual lattice site, but rather on the weights of different types of lattice sites. In other words, it attempts to capture the core physical information by using the weights and changes of different types of lattice sites.

For the three-dimensional Ising model, the number of nearest neighbors per lattice site is 6 (denoted as $n$ in the text). For a spin-up lattice site, the number of its nearest neighbors sharing the same spin can range from 0 to 6, resulting in 7 distinct categories. Similarly, for a spin-down lattice site, the same classification applies. This results in a total of 14 categories for all lattice sites. These 14 categories are then mapped to 14 corresponding network nodes, denoted by the symbol $C_{ij}$, where:
$i$ represents the spin state of the lattice site itself. In the Ising model, spin can only take two values: 1 for spin-up and 2 for spin-down.
$j$ represents the strength of nearest-neighbor interactions, with 7 possible values (1 through 7). Here, $j=1$ corresponds to cases where 0 nearest neighbors share the same spin as the central site, $j=2$ corresponds to 1 matching neighbor, and so on, up to $j=7$, which represents 6 matching neighbors. As shown in Fig.~\ref{fig11}

\begin{figure}[hptb]
	\includegraphics[width=7.0cm]{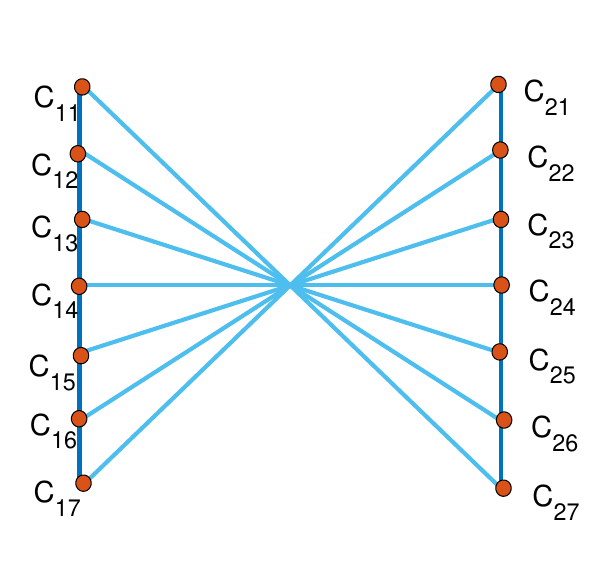}
	\caption{\label{fig11}Different network nodes represent different types of spin lattice points, while the line segments represent all possible transformation relationships.}
	\label{fig11}
\end{figure}

Now consider the flipping of a lattice site in the three-dimensional Ising model. According to the classification above, flipping a lattice site alters its category. For example, if a spin-up lattice site (with all 6 nearest neighbors sharing the same spin) is flipped, it becomes a spin-down site, and all its nearest neighbors now differ from it. This type of transformation, induced by the spin flip of the lattice site itself, is termed an active transformation in this work.
Simultaneously, when the central lattice site flips, the categories of its six neighboring sites also change—though the neighboring sites themselves remain unchanged. This occurs because the central site is a nearest neighbor to these six sites. The alteration of the central site’s state directly impacts the categories of its neighbors. This indirect transformation caused by the central site’s flip is referred to as a passive transformation.
All possible transformations are connected via edges in the network. For instance, a transition from $C_{17}$ to $C_{21}$ can occur:
$C_{17}$ corresponds to a spin-up lattice site with all 6 nearest neighbors sharing the same spin.
After flipping, the site becomes spin-down, and all its neighbors now differ from it, corresponding to $C_{21}$.

By mapping all such transformations, the network structure is constructed as Fig.~\ref{fig11}.

This network structure rigorously encompasses all possible lattice site types and transformation relationships. The weights of distinct network nodes represent the relative prevalence of each lattice site category within the original lattice model, with the total weight across all nodes summing to 1. Consequently, this network model is applicable to infinite systems, enabling its use for studying the infinite three-dimensional Ising model.

In this paper, the maximum entropy structure is used to describe the most random state of a lattice model under certain conditions. Described using the network nodes mentioned earlier, it means that the distribution of weights among different nodes is in its most random state. Knowing the weight of any single node segment for the direct calculation of the transition probabilities of all connected nodes based on factors such as temperature and interaction strength. From this, the weights of all connected nodes can be obtained using the detailed balance equation, and $\sum_{ij} C_{ij}=1$ the sum of the weights of all nodes equals 1. The network structure that satisfies the above conditions is referred to as the maximum entropy structure in this paper, denoted as $\mathbb{M}$.

This structure is directly connected to the central nodes, which refer to a class of network nodes in the Monte Carlo algorithm that can mutually transform through lattice spin flips, with all transition probabilities equal to 1. Therefore, the central nodes exhibit a direct relationship with the maximum entropy network structure. In the network model of the three-dimensional Ising transition depicted in Fig.~\ref{fig11}, the central nodes correspond to $C_{14}$ and $C_{24}$.

For the case where all spins are identical, the weight of one corresponding network node is 1, while the weights of all other network nodes are 0, and this network node possesses the lowest energy among all network nodes.
 
For instance, the ground state of the ferromagnetic Ising model, where all spins are oriented in the same direction, falls into this category. In this paper, it is referred to as the single-node structure, denoted as $\mathbb{D}$. Subsequently, the discussion will primarily focus on the phase-transition relationship between single-node structure and the maximum entropy network structure. Other ordered structures are analogous to this situation. Both structures are stable under different conditions.


This paper introduces the concept of intermediate structures $\mathbb{B}$. Specifically, there is no large-scale direct transition between the two structures mentioned above. Instead, the transition process primarily involves converting into an intermediate structure first, and then transitioning from this intermediate structure to another network structure, which is detailed later as passive transformation. Intermediate structures typically serve as the boundary between the maximum entropy network structure and the single-node structure, separating the two stable structures to form a relatively stable state. This boundary does not always exist and may only form after symmetry breaking occurs.
$\mathbb{D}$ , $\mathbb{B}$ , and $\mathbb{M}$ can all be represented by different values of $C_{ij}$.

To provide more detailed clarification, regarding the introduction of the bond concept: A bond represents the connection between two adjacent lattice points, signifying the interaction between neighboring sites. In the Ising model, this bond can take two values depending on whether the two nearest-neighbor lattice points are aligned (same spin) or misaligned (opposite spins). For ferromagnetic systems, a bond is assigned -1 (positive bond) when neighboring spins are aligned, and +1 (negative bond) when they are misaligned. This indicates a tendency for neighboring spins to align. As temperature increases, the relative weights of these two bond types change: the weight $P_1$ of lower-energy bonds (aligned spins) gradually decreases, while the weight $P_2$ of higher-energy bonds (misaligned spins) increases.
When analyzing network nodes based on bond types, each lattice point is associated with $n$ bonds (where $n/2$ is the dimension), with varying numbers of positive and negative bonds. Node weights can thus be calculated through bond weight combinations: a single-node structure corresponds to $P_1^{n}$ , boundary structures correspond to $nP_1^{2n-1}P_2$, and so on. At extremely low temperatures, $P_1$ approaches unity while $P_2$ becomes negligible, concentrating the weight distribution on single-node configurations. As temperature rises, weights progressively shift to higher-order structures.
Considering only passive transformations (where the central spin remains fixed while surrounding spins flip), at low temperatures neighboring spins strongly favor alignment with the central spin. This alignment probability diminishes with increasing temperature, thereby altering the weight distribution patterns. The system evolves toward disorder with rising temperature, causing boundary regions to shrink. Since this study focuses on critical phenomena near phase transitions, it becomes reasonable to classify nodes adjacent to single-node structures as boundary nodes - particularly crucial in the critical region where system behavior undergoes dramatic changes. Therefore, based on the preceding discussion, the boundary structure $\mathbb{B}$ of the three-dimensional Ising model in the critical region corresponds to $C_{16}$.

For the ferromagnetic three-dimensional Ising model (Fig~\ref{fig11}) at zero temperature, all lattice sites align uniformly either spin-up or spin-down (spin-up is chosen as the example here). This uniform alignment corresponds to $C_{17}$  having a weight of 1, while all other network nodes have zero weight. After the phase transition, the weights of nodes in the first and second columns of the network become equal, each contributing 1/2 to the total weight (note: this work assumes that post-transition node weight distributions follow temperature-dependent random distributions, which can be directly computed as they are independent of the primary analysis framework; further details are omitted here).

Next, analogous to Monte Carlo methods, the transition probabilities between adjacent columns of network nodes can be calculated using detailed balance. For example, consider a spin-up lattice site ($C_{17}$) where all six nearest neighbors are also spin-up. Flipping this site transitions it to $C_{21}$, and vice versa. The detailed balance formula directly yields the weight ratio between $C_{21}$ and $C_{17}$. This calculation applies only to transitions between nodes in adjacent columns (e.g., $C_{17}\leftrightarrow C_{21}$), not to transitions within the same column. This treatment aligns entirely with the Monte Carlo approach.

Specifically, in a strongly correlated system, when the spin of a particular lattice site changes, the spin type of that lattice site also changes, resulting in a change in the weight of the network node to which this lattice site belongs. This type of transformation is referred to as an active transformation in this paper. Meanwhile, the network nodes associated with the neighboring lattice sites of this lattice site also undergo changes, and this type of transformation is termed a passive transformation in this context. The active transformation corresponds to a change in $i$ within $C_{ij}$, while the passive transformation corresponds to a change in $j$.

When updating a lattice model by flipping a lattice site, it can be described from two perspectives: actively as $\partial_i C_{ij}$ or passively as $\partial_j C_{ij}$, with both descriptions being equivalent. If each lattice site has $n$ interacting neighboring sites, then the following formula can be obtained.
\begin{equation}
  \partial_i C_{ij} =n \partial_j C_{ij}
 \label{Eq:3}
\end{equation}

In the three-dimensional Ising model, flipping a single lattice site induces changes that can be represented by edges in the network. For example, $C_{17} \leftrightarrow C_{21}$ describes a successful spin flip via an active transformation. Simultaneously, the same physical flip can be interpreted as a passive transformation: when the central site flips, its six neighboring sites passively transition from $C_{17}$ to $C_{16}$ (since their shared spin alignment with the central site changes). Both descriptions correspond to the same physical flip, but differ in focus:
Active transformation: The central site itself changes state ($C_{17} \leftrightarrow C_{21}$).
Passive transformation: Six neighboring sites change state ($C_{17} \leftrightarrow C_{16}$).
Crucially, the number of passive transformations triggered by a single flip is $n$-fold greater than the active transformation, where $n$ is the number of nearest neighbors (6 in 3D). This reflects the combinatorial impact of a single spin flip on its surrounding lattice sites.

Next, I investigate the phase transition based on these distinctions. For the three-dimensional Ising model, there are two critical types of network nodes:
1. $C_{17}$  At zero temperature, all lattice sites are spin-up, corresponding to $C_{17}$ having a weight of 1, while all other nodes have zero weight.
2. $C_{14}$ and $C_{24}$: By classification rules, these nodes represent configurations where the number of spin-up and spin-down nearest neighbors are equal. Using the detailed balance equation, the weights of $C_{14}$ and $C_{24}$ are found to be equal, making them central nodes. These three node types form the basis for analyzing phase transitions.

As temperature increases, weight flows from $C_{17}$ to $C_{16}$, then splits into $C_{15}$ and $C_{14}$.
At $C_{14}$, half the weight transfers to $C_{24}$, triggering symmetry breaking and the phase transition.
Metaphorical Explanation: Imagine a flock of sheep attempting to cross a river. Most sheep cannot ford the deep channel directly (analogous to $C_{17} \leftrightarrow C_{21}$) but instead follow the shallow banks (represented by $C_{16}$-mediated pathways). As temperature rises, the "flow" of sheep shifts from deep to shallow routes, culminating in a split at the critical point ($C_{14} \leftrightarrow C_{24}$).
Role of Boundary Structure $C_{16}$:
The transition relies on $C_{16}$ acting as a boundary structure that mediates between stable states. Unlike high-dimensional Ising models where direct state conversion is inefficient,$C_{16}$ persists as a transient hub connecting higher-layer nodes (e.g.,$C_{17}$) to lower-layer nodes (e.g.,$C_{15}$). Its position—adjacent to both stable and critical nodes—enables it to regulate weight redistribution during phase transitions. The necessity of $C_{16}$ for finite-size effects and its structural linkage to base network nodes will be elaborated further below.

The flipping of lattice sites is a stochastic process: throughout the system, sites continuously transition from $C_{17}$ to $C_{16}$, while a large number simultaneously transition back from $C_{16}$ to $C_{17}$, maintaining equilibrium. This equilibrium applies to all network node transitions. $C_{16}$ is classified as a boundary structure because, after sites transition en masse from $C_{17}$ to $C_{16}$, they are more likely to revert to $C_{17}$ or remain in $C_{16}$ rather than transitioning to $C_{15}$ (temporarily ignoring active transformations). Below, we explain why $C_{16}$ preferentially reverts to $C_{17}$ rather than ultimately transitioning to $C_{14}$.

The probability of active transformation can be directly calculated from the independent variables related to fluctuations. The transition probabilities remain relatively stable before and after the phase transition, making it difficult to directly observe the phase transition. However, the combination of passive and active transformations can lead to very drastic changes. Therefore, this paper focuses on passive transformation to calculate and describe the phase transition. During the process of passive transformation, both stable structures undergo intermediate structure as boundary before transitioning into the other structure. Nodes closely related to the maximum entropy network structure are referred to as central nodes. The characteristic of central nodes is that the probability of active transformation between different $i$ values is equal to 1. This allows for rapid weight distribution among different $i$ values. Following this, passive transformation leads to changes in different $j$ values, enabling rapid and comprehensive weight distribution through the central nodes. As a result, a significant portion of the weight of a central node reached through passive transformation will be converted to other nodes. Therefore, the weights of central nodes in this paper correspond to the weights of the maximum entropy network structure.

The transition of weights among different nodes induced by fluctuations satisfies the principle of least action, meaning that when a fluctuation occurs, the required energy consumption is determined, and the transition efficiency between different nodes is maximized. This simplifies and directly relates the correspondence between the fluctuation relations of different nodes. The energy consumed by a single-node lattice flip is determined, and the maximum number of boundary lattice nodes that can be transformed into is also directly obtainable.

Let's explore the correspondence of fluctuations in three types of network structures, starting with the flow of lattice site weights among different types of network nodes.

For a lattice site in a single-node network structure, considering only the nearest-neighbor scenario, when this lattice site flips, $n$ lattice sites undergo passive changes. Due to the principle of least action, these $n$ lattice sites transition from the single-node network structure to an intermediate structure. The transition from the intermediate structure to the maximum entropy structure is also accomplished through lattice site flips. Therefore, the transition rule from the intermediate structure to the maximum entropy network structure is to flip one lattice site in the intermediate structure and calculate how many other lattice sites in the intermediate structure can passively transition to central nodes corresponding to the maximum entropy network structure. This determines the transition from lattice sites in the intermediate structure to the maximum entropy network structure. If flipping one lattice site in the intermediate structure cannot result in any passively transitioning lattice sites to the maximum entropy network structure, then multiple lattice sites in the intermediate structure are flipped simultaneously to achieve the effect of passive transition to the maximum entropy network structure. In the second process, the minimum number of passive flips achieved by flipping a single intermediate lattice site is $k$, also selected based on the principle of least action. Thus, a flip in the single-node network ultimately leads to $[nk]$ lattice sites in the maximum entropy network structure being passively generated, where $[]$ denotes the floor function since the number of particles cannot be fractional. This allows us to obtain the specific correspondence of fluctuations. Specifically, a fluctuation corresponding to a randomly selected lattice site in the single-node structure corresponds to fluctuations in different numbers of lattice sites in the intermediate and maximum entropy structures, with these numbers required to be integers. 

This correspondence arises because when the number of lattice points in a specific class of network nodes increases by 1, the maximum possible increment in fluctuation count within such nodes is 1, while there remains a finite probability for the fluctuation count to remain unchanged. Near the critical region, as the number of lattice points in a given network node class increases by 1, the probability of the fluctuation count simultaneously increasing by 1 grows monotonically. These conditions collectively satisfy the correspondence relationship discussed above.

The above provides the correspondence between fluctuations, with a minimum of 1 for these fluctuations. This means that the change in the probability of a fluctuation occurring with changes in the independent variable is relatively a minimum of 1. The smallest fluctuation is an active transformation, and different fluctuations are uncorrelated and can be completed independently. That is, the number of lattice sites undergoing fluctuations increases or decreases one by one, while passive transformation fluctuations are derived from active transformation fluctuations.

In this paper, I do not consider the cases of vacancies and double occupations, so fluctuations correspond to flips. The above sections have provided a detailed introduction on how to transition from $\mathbb{D}$ to $\mathbb{M}$ through lattice site flips, with the formula involving $\partial_i$ and $ \partial_j$.
\begin{equation}
  \partial_i \mathbb{D}=n \partial_j \mathbb{D}=n \mathbb{B}
 \label{Eq:4}
\end{equation}
Here, $\mathbb{B}$ denotes the lattice points that belong to $\mathbb{B}$.
The entire formula represents flipping a lattice point that belongs to $\mathbb{D}$, passively generating $n$ lattice points that belong to $\mathbb{B}$.
Similarly, $\mathbb{B}$ can be flipped to obtain $\mathbb{M}$, as shown in the following equation: flipping a lattice point that belongs to $\mathbb{B}$ passively generates $k$ lattice points that belong to $\mathbb{M}$.
\begin{equation}
  \partial_i \mathbb{B}=k \partial_j \mathbb{M}
 \label{Eq:5}
\end{equation}

For the results of Eq~\ref{Eq:4} and Eq.~\ref{Eq:5}, a detailed explanation is provided below.
$Q_1$ represents the probability of selecting a single-node structure, while $Q_2$ denotes the probability of selecting other nodes. As derived earlier, when considering only the weights of a column of nodes, the weights of different network nodes can be calculated as:

\begin{equation}
  (P_1+P_2)^n=P_1^{n}+nP_1^{n-1}P_2+...+P_2^n
 \label{Eq:51}
\end{equation}

Following the same logic as Monte Carlo (MC) operations, network lattice points are randomly selected and flipped with a certain probability. Here,$Q_1$ corresponds to the probability of choosing a single-node structure, and $Q_2$ to the probability of selecting other nodes.
If a lattice point has already been flipped once, this generates $n$ passive transformations, which are randomly assigned to $Q_1$ and $Q_2$. The scenario where all transformations belong to $Q_1$ trivially satisfies the requirements. Now consider cases involving both $Q_1$ and $Q_2$. Using the formula derived earlier:

\begin{equation}
  Q_2=(P_1+P_2)^n-P_1^n=nP_1^{n-1}P_2+...+P_2^n
 \label{Eq:52}
\end{equation}

Expanding the binomial terms:

\begin{equation}
  nQ_1^{n-1}Q_2 +2(n-1)Q_1^{n-1}Q_2^{2}...
  =nQ_2(Q_1+Q_2)^{n-1}=nQ_2
 \label{Eq:53}
\end{equation}

Similarly, for the $Q_1$-dominated terms:

\begin{equation}
  nQ_1^{n}+nQ_1^{n-1}Q_2+2(n-1)Q_1^{n-1}Q_2^{2}....=nQ_1(Q_1+Q_2)^{n-1}=nQ_1
 \label{Eq:54}
\end{equation}

Summing all possible probabilities, these expressions demonstrate that the total probability for a central lattice point to undergo passive transformation into a boundary lattice point is $nQ_1$. This quantifies the dominance of $Q_1$-driven transformations at low temperatures and the gradual shift toward $Q_2$ contributions as thermal fluctuations intensify.

Therefore, the unit path for transforming $\mathbb{D}$ into $\mathbb{M}$ through flipping can be obtained, as shown in the following equation, where $[nk]$ lattice points belonging to $\mathbb{M}$ will be generated each time, with $[]$ denoting the integer part.
\begin{equation}
  \partial_i (\mathbb{D} \Rightarrow \mathbb{M})=[nk] \mathbb{M}
 \label{Eq:6}
\end{equation}

Simplifying the phase transition caused by fluctuations to changes in the weights of three network structures, and more specifically, to the flow of weights among these three different network structures. There are multiple scenarios for the transition relationships among the three network structures, but due to the eq.~\ref{Eq:4} and eq.~\ref{Eq:5}, the corresponding relationship for such transitions becomes unique. Different network structures have different step sizes in the changes caused by fluctuations. If the number of fluctuations occurring per unit time in the single-node structure is continuous during the continuous change of the independent variable, then based on the correspondence relationship between fluctuations mentioned above and the correspondence relationship between central nodes and the maximum entropy structure, the passive fluctuation step size of central nodes caused by fluctuations can be seen as $[nk]$. In other words, the change in the weight of central nodes due to passive changes can be regarded as having a step size of $[nk]$. That is, during a complete weight transfer process from the single-node structure to the maximum entropy structure, the smallest unit of increase in the weight of central nodes is $[nk]$, meaning that at least $[nk]$ lattice sites are simultaneously generated through passive transitions. Therefore, among all transitions, the weight-related changes associated with the maximum entropy structure account for a weight of $P$ among all changes with a step size of $[nk]$. This illustrates the primary pathway that dominates the passive transformation. Although other pathways may also passively generate central sites, their impact on the overall outcome is negligible and requires case-specific analysis.

For active fluctuations in single-node structures, this corresponds to the generation of $n$ boundary lattice points. This work now establish a general correspondence:
For boundary structures, the addition of one boundary lattice point can induce at most one boundary fluctuation (i.e., an increase in boundary node count by 1), which may either remain stable or undergo a flip with a certain probability. Thus, a single flip of a boundary lattice point can lead to up to $n$ boundary flips.
Meanwhile, the flipping of boundary lattice points can probabilistically generate central lattice points, thereby linking single-node structures to central configurations.
As established earlier, in the critical region, boundary structures correspond to the nearest-neighbor configurations of single-node structures within the same column. Boundary lattice points can directly interact with central lattice points, enabling direct transformation between them. Consequently, multiple boundary flips can collectively generate a central lattice point.
For the 3D Ising model, there exist two pathways:
Dominant pathway: Direct generation of central lattice points from boundary flips.
Suppressed pathway: A secondary mechanism entirely inhibited in this regime.
Impact of the suppressed pathway:
From the derived formula, it can be directly calculated that the generation of one central lattice point corresponds to 1/48 of an active fluctuation in a single-node structure. Thus, (1-1/48) of the active fluctuations correspond to the generation of 3 central lattice points. This relationship allows the determination of the critical exponent governing the phase transition behavior in the system. So $\beta=3/(1-1/48)$.

For the higher-dimensional Ising model, a single flip of an single-node structure can generate $n$ boundary sites, which corresponds to a maximum of $n$ fluctuations. From the perspective of passive transformation, the conversion of a single-node structure site into a central site requires at least $n/2$ fluctuations to occur. Consequently, a single spin flip process can passively generate at most 2 central sites. Therefore, the corresponding critical exponent is determined to be 2.

Next, this work directly calculate the probability of the transformation occurring. When actively flipping a lattice point belonging to $\mathbb{D}$, the passively transformed lattice point may belong to either $\mathbb{D}$ or $\mathbb{B}$. This work decompose the transformation from $\mathbb{D}$ to $\mathbb{M}$ into two parts: $\mathbb{D}$ to $\mathbb{B}$ and $\mathbb{B}$ to $\mathbb{M}$.

\begin{equation}
  \partial_i (\mathbb{D} \Rightarrow \mathbb{M})=\partial_i (\mathbb{D} \Rightarrow \mathbb{B}) \oplus \partial_i (\mathbb{B} \Rightarrow \mathbb{M})
 \label{Eq:7}
\end{equation}

In this context, $p$ represents the probability of transitioning to the maximum entropy structure in the absence of correlations, which can be given directly. Since the step size of the transition is $[nk]$, this work only consider the situation where $[nk]$ lattice sites change simultaneously. This is equivalent to randomly selecting $[nk]$ lattice sites, and The probability associated with transitioning to the maximum entropy structure is the likelihood of being related to it, excluding the scenario where all $[nk]$ sites are unrelated to the maximum entropy structure. sites are unrelated to the maximum entropy structure, i.e., flipping one node results in $n$ passive transitions. If these $n$ lattice sites are all unrelated to the maximum entropy structure, then the selected $n$ lattice sites belong to the single node structure or the intermediate structure. From this, the follow formula can be derived.

The probability of a complete transformation process that is unrelated to $\mathbb{M}$ is

\begin{equation}
  \partial_i (\mathbb{D} \Rightarrow \mathbb{B})=(1 - p)^{[nk]}
 \label{Eq:8}
\end{equation}
So the probability of transforming into $\mathbb{M}$ during the entire process, while being unrelated to $\mathbb{D}$, is
\begin{equation}
  \partial_i (\mathbb{B} \Rightarrow \mathbb{M})=1-(1 - p)^{[nk]}
 \label{Eq:9}
\end{equation}

The weight associated with changes involving intermediate structures is necessarily related to the intermediate structures during the transition process, so the probability is 1.

\begin{equation}
 \mathcal{T}(\Gamma^{'},\Gamma)P_{eq}(\Gamma) = \mathcal{T}(\Gamma,\Gamma^{'})P_{eq}(\Gamma^{'})  
 \label{Eq:2}
\end{equation}

Within the system, $\Gamma$ and $\Gamma^{'}$ represent distinct values, with $P_{eq}$ representing their respective weights and $\mathcal{T}$ indicating the transition probabilities between them.
From the relationship between active and passive transformations, we can derive the detailed balance relationship satisfied by $\mathbb{B}$ and $\mathbb{M}$ in passive transformations.

\begin{equation}
   (\frac{m P_{BM}}{1-P_{MB}})^{n}
    \label{Eq:10}
\end{equation}
Where $P_{BM}$ represents the transition probability from an intermediate structure to the maximum entropy structure, and $P_{MB}$ represents the probability of transitioning from the maximum entropy structure to an intermediate structure and then returning to the maximum entropy structure.

Where the numerator represents the probability of transitioning into central nodes. If there are $m$ central nodes and the transition probability between central nodes is 1, it will be multiplied by $m$. The denominator represents the probability of transitioning from central nodes to the intermediate structure, which is $1$, but there is a certain probability of returning during the transition to the intermediate structure, leading to the form of the denominator.

However, due to the presence of strong correlations, during the transition between the intermediate structure and the maximum entropy structure, situations where only a single lattice site changes do not exist. Only situations where $n$ lattice sites change simultaneously are present. In other words, when only considering the maximum entropy structure and the intermediate structure, one flip corresponds to the mutual conversion of $n$ intermediate structures and maximum entropy structures. There is no situation where a single lattice site is passively converted into an intermediate structure or a maximum entropy structure. The minimum step size of the existing transition is $n$. This means that $n$ lattice sites of the intermediate structure simultaneously convert into the maximum entropy structure, or vice versa.

Therefore, the transition probability of the detailed balance relationship is:

\begin{equation}
   1-(1-p)^{[nk]}= (\frac{m P_{BM}}{1-P_{MB}})^{n}
    \label{Eq:11}
\end{equation}

This allows us to obtain the complete higher-order detailed balance relationship.

This is a higher-order detailed balance relationship, which differs from the traditional detailed balance relationship. The traditional detailed balance relationship involves the conversion of individual particles, with the minimum unit being $1$. However, the presence of strong correlations complicates this situation, as there are at least $n$ particles changing simultaneously in each transition. When calculating the weights, multiple particles changing simultaneously are also considered, and the calculation of weights for different network structures also involves multiple particles changing simultaneously. The process has changed from originally drawing one particle from the sample each time to now drawing $n1$ particles simultaneously from the sample each time, where $n1$ is a constant. Using classical probability, the detailed balance relationship for this situation can be easily obtained.

\section{Results and discussion}
In Fig~\ref{fig1}(a), $C_{ij}$ represents all possible classification and transformation relationships. Different columns signify distinct spins, while different rows indicate varying interaction intensities. Specifically, $C_{15}$ denotes the single-node network structure, $C_{14}$ represents the intermediate structure, and $C_{13}$ and $C_{23}$ represent central nodes. It is readily apparent from the figure that the transformation from the intermediate structure to the maximum entropy structure is $C_{14}\to C_{13}$, while the transition from the maximum structure to the intermediate structure is $C_{13}\to C_{14}$ minus $C_{14}\to C_{22}$. Therefore, specific transformation formulas can be derived. In Fig~\ref{fig1}(b), for a particular spin that is identical to the spin of the single-node network structure, the central node, intermediate structure, and single-node network structure of the Ising model in different dimensions can be obtained. For the Ising model in various dimensions, the corresponding fluctuation relationships can be derived and subsequently substituted into the aforementioned higher-order detailed balance relationships to obtain the formulas describing the variation of magnetic induction intensity with temperature for the Ising model in different dimensions.

\begin{figure}[hptb]
\begin{tikzpicture}
\scope[nodes={inner sep=4,outer sep=4}]
\node[anchor=south east] (a)
  {\includegraphics[width=4cm]{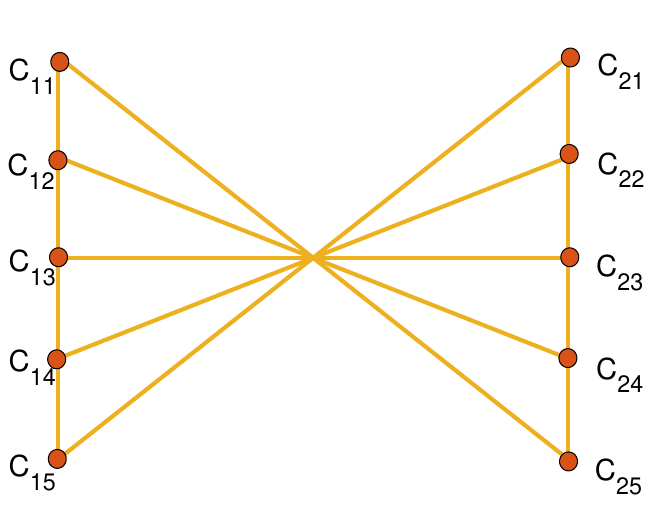}};
\node[anchor=south west] (b)
  {\includegraphics[width=4cm]{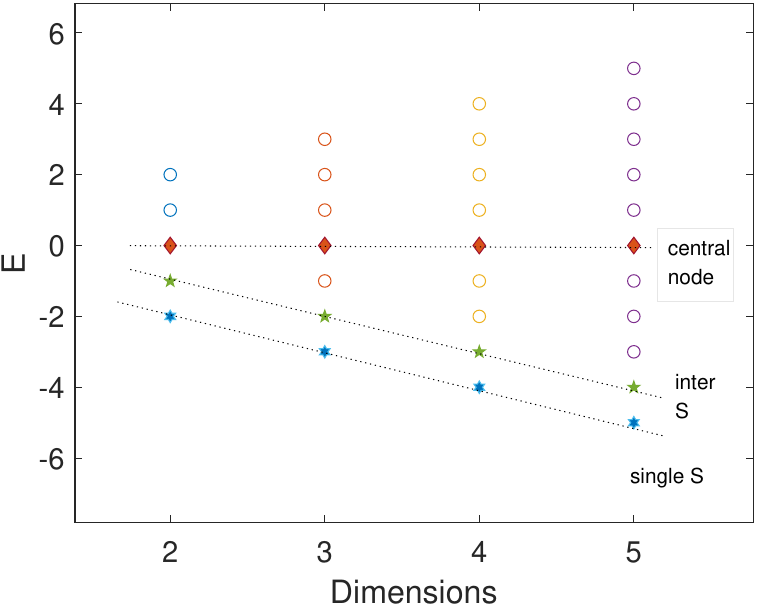}};
\endscope
\foreach \n in {a,b} {
  \node[anchor=north west] at (\n.north west) {(\n)};
}
\end{tikzpicture}
\caption{\label{fig1} (a) represents the different nodes and the transformation relationships between them in a two-dimensional Ising model, where different columns signify distinct spins, and different rows from bottom to top indicate increasing interaction strengths. (b) represents, for a specific spin, the single-node structure, intermediate structure, and central node in Ising models of different dimensions, denoted respectively by a rhombus, a pentagram, and a hexagon.}
\label{fig1}
\end{figure}

\begin{figure}[hptb]
	\includegraphics[width=7.0cm]{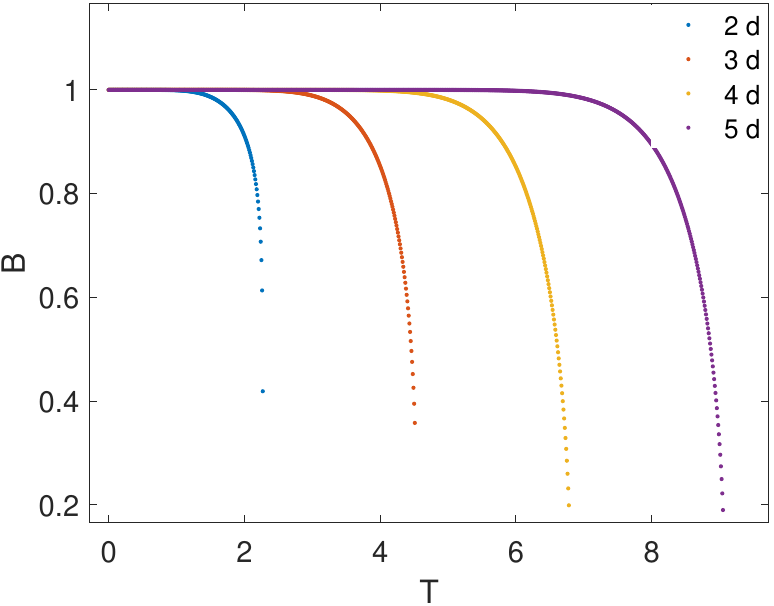}
	\caption{\label{fig2}The lines of different colors in the graph respectively represent the variation of magnetic induction intensity with temperature in two-dimensional, three-dimensional, four-dimensional, and five-dimensional Ising models.}
	\label{fig2}
\end{figure}

\begin{equation}
   1-(1-p)^{[nk]}=(\frac{2}{e^{(n-2)/T}-e^{-(n-2)/T}})^{n}
\end{equation}

\begin{equation}
        \langle m \rangle=(1-\frac{1}{{sinh}^{n}((n-2)/T)})^{1/[nk]},T\le T_c 
\end{equation} 
\begin{equation}
    \langle m \rangle=0,T\ge T_c
\end{equation}

As shown in Fig~\ref{fig2}, by substituting the fluctuation and transformation relationships of the Ising models from two to five dimensions, we can obtain the critical exponents as 1/8, 1/3, 1/2, 1/2, respectively, along with the phase transition points.

\begin{equation}
   1-(1-2p)^{3}=(\frac{2}{e^{4/T}-e^{-4/T}})^6 
   \label{eq:18}
\end{equation}

\begin{equation}
        \langle m \rangle=(1-\frac{1}{{sinh}^6(4/T)})^{1/3},T\le T_c 
\end{equation} 
\begin{equation}
    \langle m \rangle=0,T\ge T_c
\end{equation}

As shown in Fig.~\ref{fig11}, $C_{16}$ can be directly converted to $C_{14}$, and $C_{15}$ can also be converted to $C_{14}$. The conversion pathway from $C_{16}$ to $C_{14}$ dominates, allowing the critical exponent $\beta$ to be directly calculated as 1/3. If the influence of $C_{15} \rightarrow C_{14}$ conversion is considered—by evaluating the weight of this pathway through the effect of increasing the number of units in $C_{14}$ on the flipping of $C_{17}$-the original result for $\beta$ is corrected to (1-1/48)/3.

For the three-dimensional Ising model, $C_{16}$ acts as a boundary point. The conversion from $C_{16}$ to $C_{15}$ (with a conversion probability 
$e^{-2/T}$), followed by the $C_{15} \rightarrow C_{14}$ conversion (analogous to Eq.~\ref{Eq:10}), yields a relation of 
$2/(e^{2/T}-e^{-2/T})$. Consequently, Eq.~\ref{eq:18} can be modified to the following expression:

\begin{equation}
   1-(1-2p)^{3/(1-1/48)}=(\frac{47}{48}\frac{2}{e^{4/T}-e^{-4/T}}+\frac{e^{-2/T}}{48}\frac{2}{e^{2/T}-e^{-2/T}})^6  
   \label{eq:21}
\end{equation}

The discrepancy between this result and the most precise numerical values of the critical exponent $\beta$ (0.32641871) currently available is less than $0.01\%$. Given the high computational precision of Monte Carlo and renormalization algorithms, and taking into account the effects of periodic boundary conditions, the direct calculation of the critical exponent $\beta$ in this study evidently yields a highly reasonable result. Through the Eq.~\ref{eq:21}, the phase transition point of the three-dimensional Ising model is determined to be approximately 4.51, achieving high precision as well.



\section{Conclusion}
\label{conclusion}
Addressing the issue that existing theories do not clearly describe how fluctuations lead to order-disorder phase transitions, this paper employs a method of networking lattice models and the principle of maximum entropy to classify phases under different independent variables into single-node network structures, maximum entropy network structures, and intermediate structures, which correspond to ordered phases, disordered phases, and the boundaries between the two phases, respectively. The process of phase transitions is mapped onto changes in the weights of these three network structures, and the flow of weights between network nodes clearly delineates this process.

The traditional detailed balance equation may not necessarily apply in strongly correlated systems because transitions between states for particles are not completed one by one but rather by multiple particles simultaneously. Furthermore, there exists a correspondence between fluctuations occurring in different phases, naturally requiring that the slowest fluctuation has a step size of 1, so that fluctuations in other phases, derived from this correspondence, are integer multiples of the slowest fluctuation. Based on these conclusions and classical probability, a high-order detailed balance relationship similar to continuity equations is derived.

Finally, using the high-order detailed balance relationship, the critical exponents and phase transition points of the Ising model in different dimensions are given, and the critical exponents for the 2D to 5D Ising models are specifically calculated as 1/8, (1-1/48)/3, 1/2, and 1/2, respectively. Furthermore, the rationality of the theory presented in this paper is further verified through two additional problems: the antiferromagnetic frustration scenario on a triangular lattice and the estimation of the ground state energy of the two-dimensional Edwards-Anderson model.

\section{Acknowledgments}
\label{acknowledgments}
This paper is supported by the National Natural Science Foundation of China-China Academy of Engineering Physics(CAEP)Joint Fund NSAF(No. U2230402).

\nocite{*}

\bibliography{newp}

\begin{thebibliography}{33}%
\makeatletter
\providecommand \@ifxundefined [1]{%
 \@ifx{#1\undefined}
}%
\providecommand \@ifnum [1]{%
 \ifnum #1\expandafter \@firstoftwo
 \else \expandafter \@secondoftwo
 \fi
}%
\providecommand \@ifx [1]{%
 \ifx #1\expandafter \@firstoftwo
 \else \expandafter \@secondoftwo
 \fi
}%
\providecommand \natexlab [1]{#1}%
\providecommand \enquote  [1]{``#1''}%
\providecommand \bibnamefont  [1]{#1}%
\providecommand \bibfnamefont [1]{#1}%
\providecommand \citenamefont [1]{#1}%
\providecommand \href@noop [0]{\@secondoftwo}%
\providecommand \href [0]{\begingroup \@sanitize@url \@href}%
\providecommand \@href[1]{\@@startlink{#1}\@@href}%
\providecommand \@@href[1]{\endgroup#1\@@endlink}%
\providecommand \@sanitize@url [0]{\catcode `\\12\catcode `\$12\catcode `\&12\catcode `\#12\catcode `\^12\catcode `\_12\catcode `\%12\relax}%
\providecommand \@@startlink[1]{}%
\providecommand \@@endlink[0]{}%
\providecommand \url  [0]{\begingroup\@sanitize@url \@url }%
\providecommand \@url [1]{\endgroup\@href {#1}{\urlprefix }}%
\providecommand \urlprefix  [0]{URL }%
\providecommand \Eprint [0]{\href }%
\providecommand \doibase [0]{https://doi.org/}%
\providecommand \selectlanguage [0]{\@gobble}%
\providecommand \bibinfo  [0]{\@secondoftwo}%
\providecommand \bibfield  [0]{\@secondoftwo}%
\providecommand \translation [1]{[#1]}%
\providecommand \BibitemOpen [0]{}%
\providecommand \bibitemStop [0]{}%
\providecommand \bibitemNoStop [0]{.\EOS\space}%
\providecommand \EOS [0]{\spacefactor3000\relax}%
\providecommand \BibitemShut  [1]{\csname bibitem#1\endcsname}%
\let\auto@bib@innerbib\@empty
\bibitem [{\citenamefont {Luo}\ \emph {et~al.}(2022)\citenamefont {Luo}, \citenamefont {Keshtegar}, \citenamefont {Zhu}, \citenamefont {Taylan},\ and\ \citenamefont {Niu}}]{luo2022hybrid}%
  \BibitemOpen
  \bibfield  {author} {\bibinfo {author} {\bibfnamefont {C.}~\bibnamefont {Luo}}, \bibinfo {author} {\bibfnamefont {B.}~\bibnamefont {Keshtegar}}, \bibinfo {author} {\bibfnamefont {S.~P.}\ \bibnamefont {Zhu}}, \bibinfo {author} {\bibfnamefont {O.}~\bibnamefont {Taylan}},\ and\ \bibinfo {author} {\bibfnamefont {X.-P.}\ \bibnamefont {Niu}},\ }\bibfield  {title} {\bibinfo {title} {Hybrid enhanced monte carlo simulation coupled with advanced machine learning approach for accurate and efficient structural reliability analysis},\ }\href {https://doi.org/https://doi.org/10.1016/j.cma.2021.114218} {\bibfield  {journal} {\bibinfo  {journal} {Computer Methods in Applied Mechanics and Engineering}\ }\textbf {\bibinfo {volume} {388}},\ \bibinfo {pages} {114218} (\bibinfo {year} {2022})}\BibitemShut {NoStop}%
\bibitem [{\citenamefont {Prokof’ev}\ and\ \citenamefont {Svistunov}(2008)}]{prokof2008fermi}%
  \BibitemOpen
  \bibfield  {author} {\bibinfo {author} {\bibfnamefont {N.}~\bibnamefont {Prokof’ev}}\ and\ \bibinfo {author} {\bibfnamefont {B.}~\bibnamefont {Svistunov}},\ }\bibfield  {title} {\bibinfo {title} {Fermi-polaron problem: Diagrammatic monte carlo method for divergent sign-alternating series},\ }\href {https://doi.org/https://doi.org/10.1103/PhysRevB.77.020408} {\bibfield  {journal} {\bibinfo  {journal} {Physical Review B}\ }\textbf {\bibinfo {volume} {77}},\ \bibinfo {pages} {020408(R)} (\bibinfo {year} {2008})}\BibitemShut {NoStop}%
\bibitem [{\citenamefont {Foulkes}\ \emph {et~al.}(2001)\citenamefont {Foulkes}, \citenamefont {Mitas}, \citenamefont {Needs},\ and\ \citenamefont {Rajagopal}}]{RevModPhys.73.33}%
  \BibitemOpen
  \bibfield  {author} {\bibinfo {author} {\bibfnamefont {W.~M.~C.}\ \bibnamefont {Foulkes}}, \bibinfo {author} {\bibfnamefont {L.}~\bibnamefont {Mitas}}, \bibinfo {author} {\bibfnamefont {R.~J.}\ \bibnamefont {Needs}},\ and\ \bibinfo {author} {\bibfnamefont {G.}~\bibnamefont {Rajagopal}},\ }\bibfield  {title} {\bibinfo {title} {Quantum monte carlo simulations of solids},\ }\href {https://doi.org/10.1103/RevModPhys.73.33} {\bibfield  {journal} {\bibinfo  {journal} {Rev. Mod. Phys.}\ }\textbf {\bibinfo {volume} {73}},\ \bibinfo {pages} {33} (\bibinfo {year} {2001})}\BibitemShut {NoStop}%
\bibitem [{\citenamefont {Moutenet}\ \emph {et~al.}(2018)\citenamefont {Moutenet}, \citenamefont {Wu},\ and\ \citenamefont {Ferrero}}]{moutenet2018determinant}%
  \BibitemOpen
  \bibfield  {author} {\bibinfo {author} {\bibfnamefont {A.}~\bibnamefont {Moutenet}}, \bibinfo {author} {\bibfnamefont {W.}~\bibnamefont {Wu}},\ and\ \bibinfo {author} {\bibfnamefont {M.}~\bibnamefont {Ferrero}},\ }\bibfield  {title} {\bibinfo {title} {Determinant monte carlo algorithms for dynamical quantities in fermionic systems},\ }\href {https://doi.org/https://doi.org/10.1103/PhysRevB.97.085117} {\bibfield  {journal} {\bibinfo  {journal} {Physical Review B}\ }\textbf {\bibinfo {volume} {97}},\ \bibinfo {pages} {085117} (\bibinfo {year} {2018})}\BibitemShut {NoStop}%
\bibitem [{\citenamefont {Rubtsov}\ \emph {et~al.}(2005)\citenamefont {Rubtsov}, \citenamefont {Savkin},\ and\ \citenamefont {Lichtenstein}}]{rubtsov2005continuous}%
  \BibitemOpen
  \bibfield  {author} {\bibinfo {author} {\bibfnamefont {A.~N.}\ \bibnamefont {Rubtsov}}, \bibinfo {author} {\bibfnamefont {V.~V.}\ \bibnamefont {Savkin}},\ and\ \bibinfo {author} {\bibfnamefont {A.~I.}\ \bibnamefont {Lichtenstein}},\ }\bibfield  {title} {\bibinfo {title} {Continuous-time quantum monte carlo method for fermions},\ }\href {https://doi.org/https://doi.org/10.1103/PhysRevB.72.035122} {\bibfield  {journal} {\bibinfo  {journal} {Physical Review B}\ }\textbf {\bibinfo {volume} {72}},\ \bibinfo {pages} {035122} (\bibinfo {year} {2005})}\BibitemShut {NoStop}%
\bibitem [{\citenamefont {Or{\'u}s}(2019)}]{orus2019tensor}%
  \BibitemOpen
  \bibfield  {author} {\bibinfo {author} {\bibfnamefont {R.}~\bibnamefont {Or{\'u}s}},\ }\bibfield  {title} {\bibinfo {title} {Tensor networks for complex quantum systems},\ }\href {https://doi.org/https://ui.adsabs.harvard.edu/link_gateway/2019NatRP...1..538O/doi:10.1038/s42254-019-0086-7} {\bibfield  {journal} {\bibinfo  {journal} {Nature Reviews Physics}\ }\textbf {\bibinfo {volume} {1}},\ \bibinfo {pages} {538} (\bibinfo {year} {2019})}\BibitemShut {NoStop}%
\bibitem [{\citenamefont {Gray}\ and\ \citenamefont {Kourtis}(2021)}]{gray2021hyper}%
  \BibitemOpen
  \bibfield  {author} {\bibinfo {author} {\bibfnamefont {J.}~\bibnamefont {Gray}}\ and\ \bibinfo {author} {\bibfnamefont {S.}~\bibnamefont {Kourtis}},\ }\bibfield  {title} {\bibinfo {title} {Hyper-optimized tensor network contraction},\ }\href {https://doi.org/https://doi.org/10.22331/q-2021-03-15-410} {\bibfield  {journal} {\bibinfo  {journal} {Quantum}\ }\textbf {\bibinfo {volume} {5}},\ \bibinfo {pages} {410} (\bibinfo {year} {2021})}\BibitemShut {NoStop}%
\bibitem [{\citenamefont {Yuan}\ \emph {et~al.}(2021)\citenamefont {Yuan}, \citenamefont {Sun}, \citenamefont {Liu}, \citenamefont {Zhao},\ and\ \citenamefont {Zhou}}]{yuan2021quantum}%
  \BibitemOpen
  \bibfield  {author} {\bibinfo {author} {\bibfnamefont {X.}~\bibnamefont {Yuan}}, \bibinfo {author} {\bibfnamefont {J.}~\bibnamefont {Sun}}, \bibinfo {author} {\bibfnamefont {J.}~\bibnamefont {Liu}}, \bibinfo {author} {\bibfnamefont {Q.}~\bibnamefont {Zhao}},\ and\ \bibinfo {author} {\bibfnamefont {Y.}~\bibnamefont {Zhou}},\ }\bibfield  {title} {\bibinfo {title} {Quantum simulation with hybrid tensor networks},\ }\href {https://doi.org/https://doi.org/10.1103/PhysRevLett.127.040501} {\bibfield  {journal} {\bibinfo  {journal} {Physical Review Letters}\ }\textbf {\bibinfo {volume} {127}},\ \bibinfo {pages} {040501} (\bibinfo {year} {2021})}\BibitemShut {NoStop}%
\bibitem [{\citenamefont {Pan}\ and\ \citenamefont {Zhang}(2022)}]{pan2022simulation}%
  \BibitemOpen
  \bibfield  {author} {\bibinfo {author} {\bibfnamefont {F.}~\bibnamefont {Pan}}\ and\ \bibinfo {author} {\bibfnamefont {P.}~\bibnamefont {Zhang}},\ }\bibfield  {title} {\bibinfo {title} {Simulation of quantum circuits using the big-batch tensor network method},\ }\href {https://doi.org/https://doi.org/10.1103/PhysRevLett.128.030501} {\bibfield  {journal} {\bibinfo  {journal} {Physical Review Letters}\ }\textbf {\bibinfo {volume} {128}},\ \bibinfo {pages} {030501} (\bibinfo {year} {2022})}\BibitemShut {NoStop}%
\bibitem [{\citenamefont {Hu}\ \emph {et~al.}(2019)\citenamefont {Hu}, \citenamefont {Zhu}, \citenamefont {Eggert},\ and\ \citenamefont {He}}]{hu2019dirac}%
  \BibitemOpen
  \bibfield  {author} {\bibinfo {author} {\bibfnamefont {S.}~\bibnamefont {Hu}}, \bibinfo {author} {\bibfnamefont {W.}~\bibnamefont {Zhu}}, \bibinfo {author} {\bibfnamefont {S.}~\bibnamefont {Eggert}},\ and\ \bibinfo {author} {\bibfnamefont {Y.-C.}\ \bibnamefont {He}},\ }\bibfield  {title} {\bibinfo {title} {Dirac spin liquid on the spin-1/2 triangular heisenberg antiferromagnet},\ }\href {https://link.aps.org/doi/10.1103/PhysRevLett.123.207203} {\bibfield  {journal} {\bibinfo  {journal} {Physical review letters}\ }\textbf {\bibinfo {volume} {123}},\ \bibinfo {pages} {207203} (\bibinfo {year} {2019})}\BibitemShut {NoStop}%
\bibitem [{\citenamefont {Shao}\ \emph {et~al.}(2016)\citenamefont {Shao}, \citenamefont {Guo},\ and\ \citenamefont {Sandvik}}]{shao2016quantum}%
  \BibitemOpen
  \bibfield  {author} {\bibinfo {author} {\bibfnamefont {H.}~\bibnamefont {Shao}}, \bibinfo {author} {\bibfnamefont {W.}~\bibnamefont {Guo}},\ and\ \bibinfo {author} {\bibfnamefont {A.~W.}\ \bibnamefont {Sandvik}},\ }\bibfield  {title} {\bibinfo {title} {Quantum criticality with two length scales},\ }\href {https://doi.org/https://doi.org/10.1126/science.aad5007} {\bibfield  {journal} {\bibinfo  {journal} {Science}\ }\textbf {\bibinfo {volume} {352}},\ \bibinfo {pages} {213} (\bibinfo {year} {2016})}\BibitemShut {NoStop}%
\bibitem [{\citenamefont {Halperin}(2019)}]{halperin2019theory}%
  \BibitemOpen
  \bibfield  {author} {\bibinfo {author} {\bibfnamefont {B.~I.}\ \bibnamefont {Halperin}},\ }\bibfield  {title} {\bibinfo {title} {Theory of dynamic critical phenomena},\ }\href {https://doi.org/https://doi.org/10.1063/PT.3.4137} {\bibfield  {journal} {\bibinfo  {journal} {Physics Today}\ }\textbf {\bibinfo {volume} {72}},\ \bibinfo {pages} {42} (\bibinfo {year} {2019})}\BibitemShut {NoStop}%
\bibitem [{\citenamefont {Manipatruni}\ \emph {et~al.}(2019)\citenamefont {Manipatruni}, \citenamefont {Nikonov}, \citenamefont {Lin}, \citenamefont {Gosavi}, \citenamefont {Liu}, \citenamefont {Prasad}, \citenamefont {Huang}, \citenamefont {Bonturim}, \citenamefont {Ramesh},\ and\ \citenamefont {Young}}]{manipatruni2019scalable}%
  \BibitemOpen
  \bibfield  {author} {\bibinfo {author} {\bibfnamefont {S.}~\bibnamefont {Manipatruni}}, \bibinfo {author} {\bibfnamefont {D.~E.}\ \bibnamefont {Nikonov}}, \bibinfo {author} {\bibfnamefont {C.-C.}\ \bibnamefont {Lin}}, \bibinfo {author} {\bibfnamefont {T.~A.}\ \bibnamefont {Gosavi}}, \bibinfo {author} {\bibfnamefont {H.}~\bibnamefont {Liu}}, \bibinfo {author} {\bibfnamefont {B.}~\bibnamefont {Prasad}}, \bibinfo {author} {\bibfnamefont {Y.-L.}\ \bibnamefont {Huang}}, \bibinfo {author} {\bibfnamefont {E.}~\bibnamefont {Bonturim}}, \bibinfo {author} {\bibfnamefont {R.}~\bibnamefont {Ramesh}},\ and\ \bibinfo {author} {\bibfnamefont {I.~A.}\ \bibnamefont {Young}},\ }\bibfield  {title} {\bibinfo {title} {Scalable energy-efficient magnetoelectric spin--orbit logic},\ }\href {https://doi.org/https://doi.org/10.1038/s41586-018-0770-2} {\bibfield  {journal} {\bibinfo  {journal} {Nature}\ }\textbf {\bibinfo {volume} {565}},\ \bibinfo {pages} {35} (\bibinfo {year} {2019})}\BibitemShut {NoStop}%
\bibitem [{\citenamefont {Ortega-Zamorano}\ \emph {et~al.}(2015)\citenamefont {Ortega-Zamorano}, \citenamefont {Montemurro}, \citenamefont {Cannas}, \citenamefont {Jerez},\ and\ \citenamefont {Franco}}]{ortega2015fpga}%
  \BibitemOpen
  \bibfield  {author} {\bibinfo {author} {\bibfnamefont {F.}~\bibnamefont {Ortega-Zamorano}}, \bibinfo {author} {\bibfnamefont {M.~A.}\ \bibnamefont {Montemurro}}, \bibinfo {author} {\bibfnamefont {S.~A.}\ \bibnamefont {Cannas}}, \bibinfo {author} {\bibfnamefont {J.~M.}\ \bibnamefont {Jerez}},\ and\ \bibinfo {author} {\bibfnamefont {L.}~\bibnamefont {Franco}},\ }\bibfield  {title} {\bibinfo {title} {Fpga hardware acceleration of monte carlo simulations for the ising model},\ }\href {https://doi.org/https://doi.org/10.1109/TPDS.2015.2505725} {\bibfield  {journal} {\bibinfo  {journal} {IEEE Transactions on Parallel and Distributed Systems}\ }\textbf {\bibinfo {volume} {27}},\ \bibinfo {pages} {2618} (\bibinfo {year} {2015})}\BibitemShut {NoStop}%
\bibitem [{\citenamefont {Yang}\ \emph {et~al.}(2019)\citenamefont {Yang}, \citenamefont {Chen}, \citenamefont {Roumpos}, \citenamefont {Colby},\ and\ \citenamefont {Anderson}}]{yang2019high}%
  \BibitemOpen
  \bibfield  {author} {\bibinfo {author} {\bibfnamefont {K.}~\bibnamefont {Yang}}, \bibinfo {author} {\bibfnamefont {Y.-F.}\ \bibnamefont {Chen}}, \bibinfo {author} {\bibfnamefont {G.}~\bibnamefont {Roumpos}}, \bibinfo {author} {\bibfnamefont {C.}~\bibnamefont {Colby}},\ and\ \bibinfo {author} {\bibfnamefont {J.}~\bibnamefont {Anderson}},\ }\bibfield  {title} {\bibinfo {title} {High performance monte carlo simulation of ising model on tpu clusters},\ }in\ \href {https://doi.org/https://doi.org/10.1145/3295500.3356149} {\emph {\bibinfo {booktitle} {Proceedings of the International Conference for High Performance Computing, Networking, Storage and Analysis}}}\ (\bibinfo {year} {2019})\ pp.\ \bibinfo {pages} {1--15}\BibitemShut {NoStop}%
\bibitem [{\citenamefont {Preis}\ \emph {et~al.}(2009)\citenamefont {Preis}, \citenamefont {Virnau}, \citenamefont {Paul},\ and\ \citenamefont {Schneider}}]{preis2009gpu}%
  \BibitemOpen
  \bibfield  {author} {\bibinfo {author} {\bibfnamefont {T.}~\bibnamefont {Preis}}, \bibinfo {author} {\bibfnamefont {P.}~\bibnamefont {Virnau}}, \bibinfo {author} {\bibfnamefont {W.}~\bibnamefont {Paul}},\ and\ \bibinfo {author} {\bibfnamefont {J.~J.}\ \bibnamefont {Schneider}},\ }\bibfield  {title} {\bibinfo {title} {Gpu accelerated monte carlo simulation of the 2d and 3d ising model},\ }\href {https://doi.org/https://doi.org/10.1016/j.jcp.2009.03.018} {\bibfield  {journal} {\bibinfo  {journal} {Journal of Computational Physics}\ }\textbf {\bibinfo {volume} {228}},\ \bibinfo {pages} {4468} (\bibinfo {year} {2009})}\BibitemShut {NoStop}%
\bibitem [{\citenamefont {Meredith}\ \emph {et~al.}(2009)\citenamefont {Meredith}, \citenamefont {Alvarez}, \citenamefont {Maier}, \citenamefont {Schulthess},\ and\ \citenamefont {Vetter}}]{meredith2009accuracy}%
  \BibitemOpen
  \bibfield  {author} {\bibinfo {author} {\bibfnamefont {J.~S.}\ \bibnamefont {Meredith}}, \bibinfo {author} {\bibfnamefont {G.}~\bibnamefont {Alvarez}}, \bibinfo {author} {\bibfnamefont {T.~A.}\ \bibnamefont {Maier}}, \bibinfo {author} {\bibfnamefont {T.~C.}\ \bibnamefont {Schulthess}},\ and\ \bibinfo {author} {\bibfnamefont {J.~S.}\ \bibnamefont {Vetter}},\ }\bibfield  {title} {\bibinfo {title} {Accuracy and performance of graphics processors: A quantum monte carlo application case study},\ }\href {https://doi.org/https://doi.org/10.1016/j.parco.2008.12.004} {\bibfield  {journal} {\bibinfo  {journal} {Parallel Computing}\ }\textbf {\bibinfo {volume} {35}},\ \bibinfo {pages} {151} (\bibinfo {year} {2009})}\BibitemShut {NoStop}%
\bibitem [{\citenamefont {Onsager}(1944)}]{PhysRev.65.117}%
  \BibitemOpen
  \bibfield  {author} {\bibinfo {author} {\bibfnamefont {L.}~\bibnamefont {Onsager}},\ }\bibfield  {title} {\bibinfo {title} {Crystal statistics. i. a two-dimensional model with an order-disorder transition},\ }\href {https://doi.org/10.1103/PhysRev.65.117} {\bibfield  {journal} {\bibinfo  {journal} {Phys. Rev.}\ }\textbf {\bibinfo {volume} {65}},\ \bibinfo {pages} {117} (\bibinfo {year} {1944})}\BibitemShut {NoStop}%
\bibitem [{\citenamefont {Yang}(1952)}]{yang1952spontaneous}%
  \BibitemOpen
  \bibfield  {author} {\bibinfo {author} {\bibfnamefont {C.~N.}\ \bibnamefont {Yang}},\ }\bibfield  {title} {\bibinfo {title} {The spontaneous magnetization of a two-dimensional ising model},\ }\href {https://doi.org/https://doi.org/10.1103/PhysRev.85.808} {\bibfield  {journal} {\bibinfo  {journal} {Physical Review}\ }\textbf {\bibinfo {volume} {85}},\ \bibinfo {pages} {808} (\bibinfo {year} {1952})}\BibitemShut {NoStop}%
\bibitem [{\citenamefont {Ferrenberg}\ \emph {et~al.}(2018)\citenamefont {Ferrenberg}, \citenamefont {Xu},\ and\ \citenamefont {Landau}}]{PhysRevE.97.043301}%
  \BibitemOpen
  \bibfield  {author} {\bibinfo {author} {\bibfnamefont {A.~M.}\ \bibnamefont {Ferrenberg}}, \bibinfo {author} {\bibfnamefont {J.}~\bibnamefont {Xu}},\ and\ \bibinfo {author} {\bibfnamefont {D.~P.}\ \bibnamefont {Landau}},\ }\bibfield  {title} {\bibinfo {title} {Pushing the limits of monte carlo simulations for the three-dimensional ising model},\ }\href {https://doi.org/10.1103/PhysRevE.97.043301} {\bibfield  {journal} {\bibinfo  {journal} {Phys. Rev. E}\ }\textbf {\bibinfo {volume} {97}},\ \bibinfo {pages} {043301} (\bibinfo {year} {2018})}\BibitemShut {NoStop}%
\bibitem [{\citenamefont {Liu}\ \emph {et~al.}(2023)\citenamefont {Liu}, \citenamefont {Vatansever}, \citenamefont {Barkema},\ and\ \citenamefont {Fytas}}]{PhysRevE.108.034118}%
  \BibitemOpen
  \bibfield  {author} {\bibinfo {author} {\bibfnamefont {Z.}~\bibnamefont {Liu}}, \bibinfo {author} {\bibfnamefont {E.}~\bibnamefont {Vatansever}}, \bibinfo {author} {\bibfnamefont {G.~T.}\ \bibnamefont {Barkema}},\ and\ \bibinfo {author} {\bibfnamefont {N.~G.}\ \bibnamefont {Fytas}},\ }\bibfield  {title} {\bibinfo {title} {Critical dynamical behavior of the ising model},\ }\href {https://doi.org/10.1103/PhysRevE.108.034118} {\bibfield  {journal} {\bibinfo  {journal} {Phys. Rev. E}\ }\textbf {\bibinfo {volume} {108}},\ \bibinfo {pages} {034118} (\bibinfo {year} {2023})}\BibitemShut {NoStop}%
\bibitem [{\citenamefont {Vojta}(2003)}]{vojta2003quantum}%
  \BibitemOpen
  \bibfield  {author} {\bibinfo {author} {\bibfnamefont {M.}~\bibnamefont {Vojta}},\ }\bibfield  {title} {\bibinfo {title} {Quantum phase transitions},\ }\href {https://dx.doi.org/10.1088/0034-4885/66/12/R01} {\bibfield  {journal} {\bibinfo  {journal} {Reports on Progress in Physics}\ }\textbf {\bibinfo {volume} {66}},\ \bibinfo {pages} {2069} (\bibinfo {year} {2003})}\BibitemShut {NoStop}%
\bibitem [{\citenamefont {Santen}\ and\ \citenamefont {Krauth}(2000)}]{santen2000absence}%
  \BibitemOpen
  \bibfield  {author} {\bibinfo {author} {\bibfnamefont {L.}~\bibnamefont {Santen}}\ and\ \bibinfo {author} {\bibfnamefont {W.}~\bibnamefont {Krauth}},\ }\bibfield  {title} {\bibinfo {title} {Absence of thermodynamic phase transition in a model glass former},\ }\href {https://doi.org/10.1038/35014561} {\bibfield  {journal} {\bibinfo  {journal} {Nature}\ }\textbf {\bibinfo {volume} {405}},\ \bibinfo {pages} {550} (\bibinfo {year} {2000})}\BibitemShut {NoStop}%
\bibitem [{\citenamefont {Nelson}(2020)}]{nelson2020quantum}%
  \BibitemOpen
  \bibfield  {author} {\bibinfo {author} {\bibfnamefont {E.}~\bibnamefont {Nelson}},\ }\href@noop {} {\emph {\bibinfo {title} {Quantum fluctuations}}},\ Vol.~\bibinfo {volume} {16}\ (\bibinfo  {publisher} {Princeton University Press},\ \bibinfo {year} {2020})\BibitemShut {NoStop}%
\bibitem [{\citenamefont {Mishin}(2015)}]{mishin2015thermodynamic}%
  \BibitemOpen
  \bibfield  {author} {\bibinfo {author} {\bibfnamefont {Y.}~\bibnamefont {Mishin}},\ }\bibfield  {title} {\bibinfo {title} {Thermodynamic theory of equilibrium fluctuations},\ }\href {https://www.sciencedirect.com/science/article/pii/S0003491615003504} {\bibfield  {journal} {\bibinfo  {journal} {Annals of Physics}\ }\textbf {\bibinfo {volume} {363}},\ \bibinfo {pages} {48} (\bibinfo {year} {2015})}\BibitemShut {NoStop}%
\bibitem [{\citenamefont {Hohenberg}\ and\ \citenamefont {Krekhov}(2015)}]{hohenberg2015introduction}%
  \BibitemOpen
  \bibfield  {author} {\bibinfo {author} {\bibfnamefont {P.~C.}\ \bibnamefont {Hohenberg}}\ and\ \bibinfo {author} {\bibfnamefont {A.~P.}\ \bibnamefont {Krekhov}},\ }\bibfield  {title} {\bibinfo {title} {An introduction to the ginzburg--landau theory of phase transitions and nonequilibrium patterns},\ }\href {https://www.sciencedirect.com/science/article/pii/S0370157315000514} {\bibfield  {journal} {\bibinfo  {journal} {Physics Reports}\ }\textbf {\bibinfo {volume} {572}},\ \bibinfo {pages} {1} (\bibinfo {year} {2015})}\BibitemShut {NoStop}%
\bibitem [{\citenamefont {Ding}(2024{\natexlab{a}})}]{Ding202401}%
  \BibitemOpen
  \bibfield  {author} {\bibinfo {author} {\bibfnamefont {Y.}~\bibnamefont {Ding}},\ }\bibfield  {title} {\bibinfo {title} {Exploring the nexus between thermodynamic phase transitions and geometric fractals through systematic lattice point classification},\ }\href {https://doi.org/https://doi.org/10.1063/5.0204128} {\bibfield  {journal} {\bibinfo  {journal} {AIP advances}\ }\textbf {\bibinfo {volume} {14}},\ \bibinfo {pages} {085107} (\bibinfo {year} {2024}{\natexlab{a}})}\BibitemShut {NoStop}%
\bibitem [{\citenamefont {Ding}(2024{\natexlab{b}})}]{Ding202402}%
  \BibitemOpen
  \bibfield  {author} {\bibinfo {author} {\bibfnamefont {Y.}~\bibnamefont {Ding}},\ }\bibfield  {title} {\bibinfo {title} {In-depth investigation of phase transition phenomena in network models derived from lattice models},\ }\href {https://doi.org/https://doi.org/10.1063/5.0219207} {\bibfield  {journal} {\bibinfo  {journal} {AIP advances}\ }\textbf {\bibinfo {volume} {14}},\ \bibinfo {pages} {085308} (\bibinfo {year} {2024}{\natexlab{b}})}\BibitemShut {NoStop}%
\bibitem [{\citenamefont {Mark}\ \emph {et~al.}(2024)\citenamefont {Mark}, \citenamefont {Surace}, \citenamefont {Elben}, \citenamefont {Shaw}, \citenamefont {Choi}, \citenamefont {Refael}, \citenamefont {Endres},\ and\ \citenamefont {Choi}}]{PhysRevX.14.041051}%
  \BibitemOpen
  \bibfield  {author} {\bibinfo {author} {\bibfnamefont {D.~K.}\ \bibnamefont {Mark}}, \bibinfo {author} {\bibfnamefont {F.}~\bibnamefont {Surace}}, \bibinfo {author} {\bibfnamefont {A.}~\bibnamefont {Elben}}, \bibinfo {author} {\bibfnamefont {A.~L.}\ \bibnamefont {Shaw}}, \bibinfo {author} {\bibfnamefont {J.}~\bibnamefont {Choi}}, \bibinfo {author} {\bibfnamefont {G.}~\bibnamefont {Refael}}, \bibinfo {author} {\bibfnamefont {M.}~\bibnamefont {Endres}},\ and\ \bibinfo {author} {\bibfnamefont {S.}~\bibnamefont {Choi}},\ }\bibfield  {title} {\bibinfo {title} {Maximum entropy principle in deep thermalization and in hilbert-space ergodicity},\ }\href {https://doi.org/10.1103/PhysRevX.14.041051} {\bibfield  {journal} {\bibinfo  {journal} {Phys. Rev. X}\ }\textbf {\bibinfo {volume} {14}},\ \bibinfo {pages} {041051} (\bibinfo {year} {2024})}\BibitemShut {NoStop}%
\bibitem [{\citenamefont {Bortz}\ \emph {et~al.}(1975)\citenamefont {Bortz}, \citenamefont {Kalos},\ and\ \citenamefont {Lebowitz}}]{BORTZ197510}%
  \BibitemOpen
  \bibfield  {author} {\bibinfo {author} {\bibfnamefont {A.}~\bibnamefont {Bortz}}, \bibinfo {author} {\bibfnamefont {M.}~\bibnamefont {Kalos}},\ and\ \bibinfo {author} {\bibfnamefont {J.}~\bibnamefont {Lebowitz}},\ }\bibfield  {title} {\bibinfo {title} {A new algorithm for monte carlo simulation of ising spin systems},\ }\href {https://doi.org/https://doi.org/10.1016/0021-9991(75)90060-1} {\bibfield  {journal} {\bibinfo  {journal} {Journal of Computational Physics}\ }\textbf {\bibinfo {volume} {17}},\ \bibinfo {pages} {10} (\bibinfo {year} {1975})}\BibitemShut {NoStop}%
\bibitem [{\citenamefont {Liao}\ \emph {et~al.}(2017)\citenamefont {Liao}, \citenamefont {Xie}, \citenamefont {Chen}, \citenamefont {Liu}, \citenamefont {Xie}, \citenamefont {Huang}, \citenamefont {Normand},\ and\ \citenamefont {Xiang}}]{liao2017gapless}%
  \BibitemOpen
  \bibfield  {author} {\bibinfo {author} {\bibfnamefont {H.-J.}\ \bibnamefont {Liao}}, \bibinfo {author} {\bibfnamefont {Z.-Y.}\ \bibnamefont {Xie}}, \bibinfo {author} {\bibfnamefont {J.}~\bibnamefont {Chen}}, \bibinfo {author} {\bibfnamefont {Z.-Y.}\ \bibnamefont {Liu}}, \bibinfo {author} {\bibfnamefont {H.-D.}\ \bibnamefont {Xie}}, \bibinfo {author} {\bibfnamefont {R.-Z.}\ \bibnamefont {Huang}}, \bibinfo {author} {\bibfnamefont {B.}~\bibnamefont {Normand}},\ and\ \bibinfo {author} {\bibfnamefont {T.}~\bibnamefont {Xiang}},\ }\bibfield  {title} {\bibinfo {title} {Gapless spin-liquid ground state in the s= 1/2 kagome antiferromagnet},\ }\href {https://link.aps.org/doi/10.1103/PhysRevLett.118.137202} {\bibfield  {journal} {\bibinfo  {journal} {Physical review letters}\ }\textbf {\bibinfo {volume} {118}},\ \bibinfo {pages} {137202} (\bibinfo {year} {2017})}\BibitemShut {NoStop}%
\bibitem [{\citenamefont {Ferrari}\ and\ \citenamefont {Becca}(2019)}]{ferrari2019dynamical}%
  \BibitemOpen
  \bibfield  {author} {\bibinfo {author} {\bibfnamefont {F.}~\bibnamefont {Ferrari}}\ and\ \bibinfo {author} {\bibfnamefont {F.}~\bibnamefont {Becca}},\ }\bibfield  {title} {\bibinfo {title} {Dynamical structure factor of the j 1-j 2 heisenberg model on the triangular lattice: magnons, spinons, and gauge fields},\ }\href {https://link.aps.org/doi/10.1103/PhysRevX.9.031026} {\bibfield  {journal} {\bibinfo  {journal} {Physical Review X}\ }\textbf {\bibinfo {volume} {9}},\ \bibinfo {pages} {031026} (\bibinfo {year} {2019})}\BibitemShut {NoStop}%
\bibitem [{\citenamefont {Dominguez}\ \emph {et~al.}(2011)\citenamefont {Dominguez}, \citenamefont {Lage-Castellanos}, \citenamefont {Mulet}, \citenamefont {Ricci-Tersenghi},\ and\ \citenamefont {Rizzo}}]{dominguez2011characterizing}%
  \BibitemOpen
  \bibfield  {author} {\bibinfo {author} {\bibfnamefont {E.}~\bibnamefont {Dominguez}}, \bibinfo {author} {\bibfnamefont {A.}~\bibnamefont {Lage-Castellanos}}, \bibinfo {author} {\bibfnamefont {R.}~\bibnamefont {Mulet}}, \bibinfo {author} {\bibfnamefont {F.}~\bibnamefont {Ricci-Tersenghi}},\ and\ \bibinfo {author} {\bibfnamefont {T.}~\bibnamefont {Rizzo}},\ }\bibfield  {title} {\bibinfo {title} {Characterizing and improving generalized belief propagation algorithms on the 2d edwards--anderson model},\ }\href {https://dx.doi.org/10.1088/1742-5468/2011/12/P12007} {\bibfield  {journal} {\bibinfo  {journal} {Journal of Statistical Mechanics: Theory and Experiment}\ }\textbf {\bibinfo {volume} {2011}},\ \bibinfo {pages} {P12007} (\bibinfo {year} {2011})}\BibitemShut {NoStop}%
\end{thebibliography}%

\onecolumngrid
\newpage 
\newcounter{equationSM}
\newcounter{figureSM}
\newcounter{tableSM}
\newcounter{sectionSM}
\stepcounter{equationSM}
\setcounter{equation}{0}
\setcounter{figure}{0}
\setcounter{table}{0}
\setcounter{section}{0}
\makeatletter
\renewcommand{\theequation}{\textsc{sm}-\arabic{equation}}
\renewcommand{\thefigure}{\textsc{sm}-\arabic{figure}}
\renewcommand{\thetable}{\textsc{sm}-\arabic{table}}
\renewcommand{\thesection}{\textsc{sm}-\arabic{section}}

\begin{center}
  {\large{\bf Supplemental Material for\\
  ``Theory of Order-Disorder Phase Transitions Induced by Fluctuations Based on Network Models''}}
\end{center}
\begin{appendices}
\appendix

\section{A: Frustration in Antiferromagnetic Ising Model}\label{sm-A}
Firstly, the antiferromagnetic model on a triangular lattice is transformed into a network model. For simplicity, the initial consideration is limited to the cases where each lattice site has only two possible spin orientations: up and down. For lattice sites with spins up, similar to the approach in the paper, classification is based on the number of neighboring sites with different spin orientations, regardless of their sequence. Among these, the lattice site with the lowest energy is the one surrounded by six neighboring sites all with spins down. This allows all possible lattice sites with spins up to be categorized into $7$ classes, corresponding to $7$ network nodes. Similarly, for lattice sites with spins down, they can also be converted into $7$ network nodes. During the Monte Carlo updating process, all types of lattice sites that can undergo direct transitions are connected by lines. Thus, the entire antiferromagnetic triangular lattice model is transformed into a model with $14$ network nodes.

\begin{figure}[hptb]
	\includegraphics[width=7.0cm]{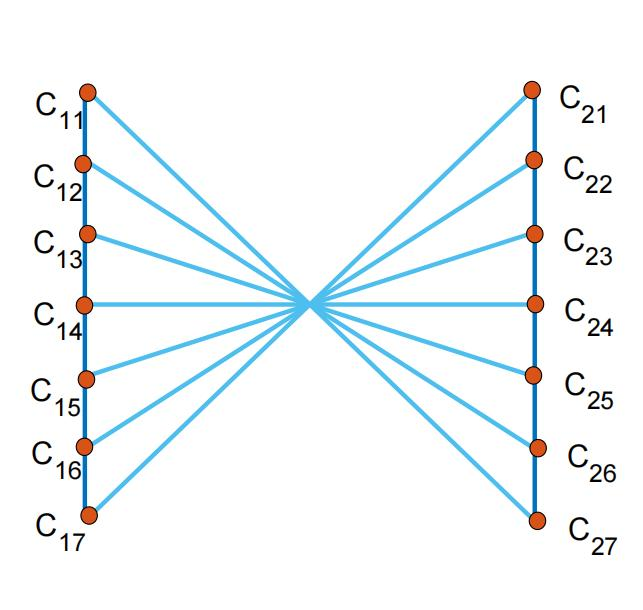}
	\caption{\label{fig:3}Different network nodes represent different types of spin lattice points, while the line segments represent all possible transformation relationships.}
	\label{fig:3}
\end{figure}

Next, I'll start with a simple scenario, assuming that the spins of different lattice sites can only be in two states: spin-up and spin-down. In the case of antiferromagnetism, for a triangular lattice, the lowest energy configuration occurs when two lattice sites have the same spin and the third site has the opposite spin. It's straightforward to observe that the lowest energy state arises when all triangles adhere to this configuration. Does such a state exist? The answer is yes. This state emerges when one row of lattice sites has spins up, followed by another row with spins down, alternating in this pattern. In the corresponding network model, this corresponds to nodes $C_{15}$ and $C_{25}$ both having weights of $0.5$, while other nodes have weights of $0$. At extremely low temperatures, this state is stable because flipping any lattice site during Monte Carlo updates would require energy. Therefore, this state is stable.

However, at higher temperatures, the system should conform to the maximum entropy model. The rods representing antiferromagnetic frustration in a triangular lattice are directly related, meaning that the three rods forming a triangle cannot simultaneously be negative. This situation differs from the Ising model in different dimensions, so classical probability methods cannot be directly used to calculate the weights of different nodes at various temperatures. Instead, we can classify all possible types of triangular lattice sites. At different temperatures, different types of triangles have different energies and thus different weights. Subsequently, we can use probability formulas to calculate the weights of different types of lattice sites at various temperatures. The blue rods indicate spins that are the same, while the yellow rods indicate spins that are opposite (note: originally, both blue and yellow rods were described as indicating the same spin, which is corrected here to reflect their distinct meanings in the context of antiferromagnetism).

\begin{figure}[hptb]
	\includegraphics[width=7.0cm]{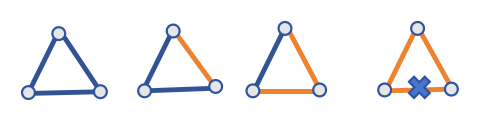}
	\caption{\label{fig:4}The blue lines represent ferromagnetic interactions, while the yellow lines represent antiferromagnetic interactions. The fourth scenario depicted in the diagram does not exist.}
	\label{fig:4}
\end{figure}

Through the above analyses, we have obtained the scenarios at extremely low temperatures and at high temperatures. It is evident that a transition from the extremely low temperature scenario to the high temperature scenario is possible. Conversely, can the high temperature scenario transition to the extremely low temperature scenario? The answer is yes. During the process of lowering the temperature, the weights of nodes with lower energy increase continuously. However, $C_{16}$ and $C_{26}$ are obviously unstable because they are inevitably surrounded by other lattice sites. In contrast, $C_{15}$ and $C_{25}$ are stable. Therefore, when the temperature drops to a certain level, the probability of transition from $C_{16}$ and $C_{26}$ to other nodes still exists. But since $C_{15}$ and $C_{25}$ are very stable, the transition from $C_{15}$ and $C_{25}$ to $C_{16}$ and $C_{26}$ is unlikely to occur. Consequently, the weights gradually concentrate on $C_{15}$ and $C_{25}$, forming the first scenario. The situations for $C_{17}$ and $C_{27}$ are similar.

Next, let's explore the transition rule from order to disorder. The central points, which are also the centers of maximum entropy, are obviously $C_{14}$ and $C_{24}$. The transition of these nodes does not result in a change in energy, and the close relationship between these nodes and the maximum entropy network is discussed in detail in the article. Since $C_{14}$ and $C_{24}$ are directly connected to $C_{15}$ and $C_{25}$, there are no boundary nodes. This situation implies that there is no phase transition point between the two phases.
Assuming each bond has a value of 1, and in this problem $n$ is 6, the formula for the weight variation with temperature in the maximum entropy network structure $p$ is given as follows.
\begin{equation}
   p=(\frac{2}{e^{(n-2)/T}-e^{-(n-2)/T}})^{n}(1-p)=(\frac{2}{e^{4/T}-e^{-4/T}})^{6}(1-p)
   \label{sm:1}
\end{equation}

\section{B: Expectation of Ground State Energy in the Two-Dimensional Edwards-Anderson Model}\label{sm-B}
The relationship between frustration and ground state energy has been established by the algorithm mentioned above. Next, we will use the number of frustrations to derive the ground state energy.

If the probability of $J=\pm 1$ taking a positive value is $q$, then for the two-dimensional Edwards-Anderson model on a square lattice, the expression for all possible combinations of $J$ values across the lattice can be written as 
\begin{equation}
    (q+(1-q))^4=q^4(1-q)^0+4q^3(1-q)^1+6q^2(1-q)^2+4q^1(1-q)^3+q^0(1-q)^4
\end{equation}

, where the expansion represents the sum of probabilities for all configurations of J

Firstly, let's consider the expectation of the ground state energy in the two-dimensional Edwards-Anderson model. In this case, the weights for both positive and negative $J$ are $0.5$. This allows us to calculate the weights of different types of lattice sites.

Each square lattice has four bonds, and there are five types of bonds, with equal weights of $0.5$ for both positive and negative $J$. From this, we can deduce the weights of different square lattices.

The condition for frustration to form is when the number of negative $J$ is either $1$ or $3$. Only in these cases will a frustrated lattice be created. Each frustrated lattice corresponds to a lattice with the lowest energy. Therefore, there must be a bond with a value of positive $1$, and this bond can be shared by two lattices. From this, we can infer the ground state energy of the entire model.

The probability $P$ for a frustrated lattice is
\begin{equation}
    P=4q^3(1-q)^1+4q^1(1-q)^3=0.5
\end{equation}

Furthermore, the lowest ground state energy can be estimated by the number of frustrated lattices is $-1\times 0.5-2\times 0.5=-1.5$.

Therefore, the weight $p$ of the maximum entropy network structure can be obtained as a function of temperature.
\begin{equation}
  1-(1-(p-0.5z))^{8}=(\frac{2}{e^{2/T}-e^{-2/T}})^{4},p-0.5z\ge 0
\end{equation}
\begin{equation}
  z=(\frac{2}{e^{(n-2)/T}-e^{-(n-2)/T}})^{n}(1-p)=(\frac{2}{e^{2/T}-e^{-2/T}})^{4}(1-z)
\end{equation}
Subtracting $0.5z$ from p is because at extremely low temperatures, the probability associated with the maximum entropy structure can also be transformed through the boundary structure, with a related probability of $0.5z$, without needing to start the transformation from a single-node structure. The relationship between the boundary structure and the maximum entropy structure satisfies Eq~\ref{sm:1}, thus leading to the above formula.

\end{appendices}

\end{document}